\documentclass[twocolumn,trackchanges]{aastex631}
\usepackage{graphicx}
\usepackage{amsmath}
\usepackage{multirow}
\usepackage{array}
\usepackage[T1]{fontenc}
\usepackage[caption=false]{subfig}
\usepackage{ulem}
\usepackage{cancel}
\usepackage{threeparttable}

\usepackage{soul}  
\usepackage{mathptmx} 
\usepackage{hyperref}  
\begin{document}

\title{Discovery of a  new flaring class in GRS 1915+105 using AstroSat}
\author{Ruchika Dhaka}
\affiliation{Department of Physics, IIT Kanpur, Kanpur, Uttar Pradesh
208016, India}

\author{J.S. Yadav}
\affiliation{Department of Physics, IIT Kanpur, Kanpur, Uttar Pradesh
208016, India}
\affiliation{Tata Institute of Fundamental Research, Homi Bhabha Road, 400005, Mumbai, India}

\author{Ranjeev Misra}
\affiliation{Inter-University Center for Astronomy and Astrophysics,
Ganeshkhind, Pune 411007, India}

\author{Pankaj Jain}
\affiliation{Department of Physics, IIT Kanpur, Kanpur, Uttar Pradesh
208016, India}

\begin{abstract}
Highly variable  Black Hole X-ray Binary (BHXB) GRS~1915+105 has shown many flaring classes when the source oscillates between the High Soft state (HS) and the Hard Intermediate state (HIMS) with a transition time of less than 10 s. The  X-ray flux is anti-correlated with hardness ratio (HR2)  during these X-ray flaring classes. We have analyzed Astrosat/LAXPC \& SXT data and report here a new X-ray flaring class named $\eta$  class when the source oscillates between two HS states (the power-law index is always greater than 4) with transition time around  50 s. The X-ray flux is correlated with hardness ratio. This class is quasi-regular, and we have detected High Frequency Quasi Periodic Oscillations (HFQPOs) around 70 Hz during this new flaring class. The accretion rate changes by a factor of three over the burst cycle.  We report here the results of our extensive study of spectral and timing characteristics of this new class.

\end{abstract}





\section{Introduction}

GRS 1915+105 is a transient BHXB comprised of a K-M iii spectral type donor star and a 12.5 solar mass black hole at a distance of 8.6 kpc \citep[][]{reid2014parallax}. This source has its relativistic jets oriented at an angle $i=70^\circ$ relative to our line of sight \citep{mirabel1994superluminal}. GRS 1915+105 contains a fast-spinning Kerr black hole with a spin around 0.99 \citep[][]{blum2009measuring, miller2013nustar, mcclintock2006spin, dhaka2023correlations, remillard2006x}. 

\citet{2022NatAs...6..577M} revealed a strong correlation between jet radio flux and disk iron-line flux, along with an anti-correlation with corona temperature, suggesting energy alternates between jet powering and corona heating, with the corona potentially transforming into the jet. \citet{2022MNRAS.513.4196G} showed a long-term evolution of the coronal size, temperature, and feedback fraction, correlating QPO frequency with disk-corona interactions and jet ejection. \citet{2024MNRAS.527.7136B} demonstrated that type-C QPO properties, including phase lags and fractional rms, evolve consistently across short and long timescales, with the corona size increasing with QPO frequency and matching parameters observed over extended periods.

GRS 1915+105  is a highly variable source and exhibits slow as well as fast variability in its light curve on a timescale of seconds to several minutes \citep[][]{1997AIPC..410..907G, belloni2000model, paul1998x}. \citet{belloni2000model} have analyzed a large sample of RXTE/PCA observations and classified them into 12 separate classes based on their lightcurve and color-color diagram (CCD).  More studies have increased this to  15 distinct classes named as $\alpha, \beta, \gamma, \delta, \theta, \kappa, \lambda, \mu, \nu, \rho, \phi, \chi, \omega, \xi$  and $\psi$ \citep[][] {belloni2000model,klein2002hard, hannikainen2005characterizing, shi2023new}. The variability in these classes may be explained as the transitions between three basic source states: a hard or HIMS state  (state C), and two softer states 
with a fully observable disk (states A and B) \citep[][] {belloni2000model}. 

All the observed classes in GRS 1915+105 can be put broadly into three categories \citep[][] {belloni2000model}:  1. Those classes that belong to the HIMS, which represents a relatively harder spectral state compared to softer states, like the $\chi$ class and its sub-classes. It may be radio loud or radio-quiet  \citep{muno2001radio, trudolyubov2001two, klein2002hard, yadav2006connection}. The $\chi$ class is seen in all X-ray binaries (XRBs) as some variant except the Plateau X-ray state, which has not been seen so far in any X-ray binaries except in  GRS 1915+105. 2. Those classes that are dominated by thermal state or HS state like  $\phi $, $\delta$, and  $\gamma$ classes. and 3. X-ray flaring classes like $ \rho, \kappa, \lambda, \theta, \beta$, and others show the fast transition between HIMS and HS  X-ray states \citep[][]{belloni2000model}. Some of these  X-ray flaring classes are radio quiet while others may be radio loud, which may also involve a fast transition between state A and state B \citep[][]{vadawale2001spectral, naik2001detection, klein2002hard}.

There has been immense interest in the X-ray flaring classes, and some of these studies were carried out even before their classification.  \citet{belloni1997unstable} have studied  $ \lambda $ flaring class. The light curve shows a complicated pattern of dips and repeated fast transitions between high flux (burst phase, which they defined as outburst) and low X-ray flux (quiescent phase of around 300 s) with a burst cycle of around 1000 s. The fast transition is termed as flaring. These are regular bursts. The X-ray flux is anti-correlated with the hardness ratio. The shortest doubling time is about 2 s while the decay time is significantly less (0.5–1 s). The fast transitions between high  X-ray flux and low X-ray flux (burst phase and quiescent phase) are explained in terms of rapid removal (disappearing) and replenishment (appearing) of the inner part of the accretion disk.  The rise time is explained as the propagation of a perturbation through the disk on the viscous timescale, while the decay time corresponds to the free-fall time of the accreting matter. \citet{belloni1997unified} have studied  $ \kappa$ X-ray flaring class using  RXTE/PCA data of June 18, 1997. These are irregular bursts with varying burst durations. Each event is defined as the sum of the quiescent duration and the burst duration. A tight correlation is observed between the quiescent duration and the following burst duration. The accretion disk radius is well correlated with the quiescent duration. All spectral changes are explained using the above model of the rapid disappearance of the inner region of an accretion disk, followed by a slower refilling of the emptied region. The mass accretion rate during the burst phase is around two times higher than that during the quiescent phase. Burst rise and decay time are explained as in the case of $ \lambda $ X-ray flaring class as described above.

\citet{yadav1999different}  have studied a large sample of $ \rho $  and  $\kappa$ classes using data of  IXAE observations during  1997 June 12 - 29 and  August 7 - 10.   All the observed bursts show slow exponential rise and sharp linear decay. These bursts are classified broadly into three groups: 
1.  regular bursts  with  low dispersion ($\delta P / P  \leq  10 \%$), 2. quasi-regular  bursts with moderate dispersion ($\delta P / P  \sim   10 - 50 \%$ and  3.  irregular  bursts with  no fixed periodicity $\delta P / P  \geq  50 \%$).
Here, P is the burst period or burst recurrence time. There is a strong correlation between the quiescent duration and the following burst duration for the quasi-regular and irregular bursts. No such 
correlation is found for the regular bursts.  All types of bursts are explained as the transition between two X-ray states of the source. 
The source can switch back and forth between the HIMS state and the HS state near critical accretion rates in a very short time, giving rise to irregular and quasi-regular bursts.  The fast time scale for the transition of the state is explained by invoking the appearance and disappearance of the advective disk in its viscous time scale.   The periodicity of the regular bursts is explained by matching the viscous time scale with the cooling time scale of the post-shock region. It is further suggested that the disappearance of the inner region of a standard thin disk on a free-fall timescale is not feasible because angular momentum transfer must occur for the material to flow inward. The time required for this transfer corresponds to the viscous timescale.

\citet{neilsen2011physics} have presented a phase-resolved spectral analysis of the $ \rho $ X-ray class in GRS 1915+105, demonstrating that the observed spectral and timing properties are consistent with the radiation pressure instability and the evolution of a local Eddington limit in the inner disk. Expanding on these findings, \citet{2017MNRAS.465.1926Y} performed a detailed timing analysis of the heartbeat state of GRS 1915+105, mapping the phase-resolved energy-frequency-power distribution and highlighting the role of aperiodic variability in tracing interactions between the corona and the accretion disk. These results corroborate the spectral findings of \citet{neilsen2011physics}, linking the slow rise phase to the local Eddington limit and the pulse rise phase to radiation pressure-driven instabilities in the disk. Additionally, \citet{2018MNRAS.474.1214Y} utilized phase-lag as a function of energy and frequency to explore disc–corona interactions, demonstrating how variability in the corona influences disc responses across different phases of the $ \rho $ state.

\citet{Mirabel:1997du}  have studied  $\beta$ class using  simultaneous observations in the X-ray,
infrared, and radio wavelengths. During episodes of rapid disappearance and the replenishment of the inner accretion disk, the ejection of relativistic plasma clouds are suggested that produce flares at infrared and radio wavelengths.  \citet{yadav2001disk} has suggested that the ``spike`` which separates the dips with hard and soft spectra in $\beta$ X-ray light curve,  marks the start of a major ejection episode of the synchrotron-emitting plasma producing infrared (IR) and radio flares.  There is a fast transition between state B and state A at the spike.  X-ray flux is anti-correlated with HR2 for hard dips \citep[][]{yadav2001disk}.  The $\beta$ and $\theta$ X-ray flare classes are both radio-loud and produce transient radio jets \citep[][]{Mirabel:1997du, dhawan2000scale}.  Both classes show hard and soft dips along with spikes in the light curve.

In this work, we use Astrosat observations from 27 July 2017 to 11 September 2017 (MJD 57960-58006), which show flaring class as listed in Table \ref{tab:table1}. Some of these data have been used in the past. \citet{belloni2019variable} have used Astrosat observations from 9 July 2017 to 12 September 2017 to study flaring and non-flaring classes and have reported HFQPOs with frequencies varying between 67.4 and 72.3 Hz. This study shows for the first time that HFQPO frequency variation and associated lags are correlated with the source position in the Hardness-Intensity Diagram (HID).  The flaring class is referred as $\omega$ class.  \citet{athulya2022} have analyzed a larger sample of  Astrosat observations from 13 November 2016 to 14 June 2019 of flaring and non-flaring classes. Using Astrosat/LAXPC data,  timing, and energy spectra are analyzed to study class variability in GRS 1915+105, but  HFQPO was not reported. The LAXPC observations during  MJD 57961-58007 are suggested to belong to $\kappa$ or $\omega$ flaring classes. \citet{majumder2022wide}  have also used a larger sample of Astrosat observations from 4 March 2016 to 23 March 2019  to study flaring and non-flaring classes in GRS 1915+105.  This study suggests that the power of HFQPO  around 70 Hz is significant during high X-ray flux regions and HFQPO is absent during the dip regions.  It is suggested that  HFQPOs are due to modulation of the Comptonizing corona.  Astrosat observations during  MJD 57691--58007 are suggested to belong to the $\kappa$ and $\omega$ classes. The $\kappa$ X-ray flaring class  has been studied in detail by 
\citet{belloni1997unified} and  \citet{yadav1999different} While $\omega$ X-ray flaring class has been studied  in detail by \citet{klein2002hard} and  \citet{naik2002}. 
During both $\kappa$ and $\omega$ X-ray flaring classes, HR2 decreases when the source moves from the dips (low flux) to the burst phase (high flux) \citep{ yadav1999different,naik2002}.

In this paper, we present our analysis results of Astrosat observation of GRS 1915+105 during July-September 2017  and report here a new X-ray flaring class when the source oscillates between two HS states (power index is always greater than 4). We have named it as the $\eta$ X-ray flaring class. The fast transition time is around  50 s, and the  X-ray flux is in phase with HR2 (it increases with X-ray flux). This class is quasi-regular, and we have detected $\sim$ 70 Hz QPO  with soft lag. The accretion rate changes by a factor of three over the burst cycle.  
In all previously known X-ray flaring classes, as discussed above, the source oscillates between HS state and HIMS with a transition time of less than 10 s, and the  X-ray flux is anticorrelated with HR2. The  $\sim$ 70 Hz QPO is never observed during any of the previously known X-ray flaring classes.  We report here the results of our extensive study of spectral and timing characteristics of this new  X-ray flaring class.

\section{Observations and data reduction}
\label{sec:obs_data_red}
AstroSat, India's first satellite dedicated to astronomical research, possesses the capability to investigate a diverse range of celestial objects across various wavelengths, including near and far ultraviolet (UV), soft X-rays (0.3 - 8 keV), and hard X-rays (3 - 100 keV)  \citep[][]{agrawal2006broad}. It has four co-aligned instruments: the Ultra-Violet Imaging Telescope (UVIT) \citep[][]{tandon2017orbit}, the Soft X-ray Telescope (SXT) \citep[][]{singh2016orbit, singh2017soft}, the Large Area X-ray Proportional Counter (LAXPC) \citep[][]{yadav2016astrosat, yadav2016large}, and the Cadmium Zinc Telluride Imager (CZTI) \citep[][]{bhalerao2017cadmium}.

For our study, we exclusively utilized the  LAXPC and SXT observations of GRS 1915+105 taken during July - September 2017. The SXT provides spectral coverage from 0.3 to 8 keV, while the LAXPC provides the advantage of a large effective area in the 3 to 80 keV range, enabling a wide-band spectral perspective. Moreover, the good temporal resolution of LAXPC, at 10 microseconds, presents an opportunity for detecting QPOs across a broad range of frequencies \citep[][]{yadav2016astrosat, belloni2019variable}.
For this work, we used a dataset consisting of a distinct observations conducted on the following dates: 27 Jul 2017 (Obs. 1), 27 Apr 2017 (Obs. 2), 30 Aug 2017 (Obs. 3), 31 Aug 2017 (Obs. 4) and 11 Sep 2017 (Obs. 5). The effective exposure times of both LAXPC and SXT, along with the corresponding AstroSat observation IDs, are presented in Table \ref{tab:table1} for reference.

\subsection{LAXPC}
Level 2 event files were extracted from Level 1 event mode data utilizing the official LAXPC software version released on 04 Aug 2020\footnote{\href{http://astrosat-ssc.iucaa.in/laxpcData}{http://astrosat-ssc.iucaa.in/laxpcData}}. LAXPC data extraction and processing was performed following \citet{antia2017calibration} to obtain the source spectra and light curves. Details of the response matrix (RMF) and background spectrum generation for proportional counters 10, 20, and 30, respectively, can be found in \citet{antia2017calibration}. The LAXPC software generates the Good Time Interval (GTI) for the data, encompassing the observation's time information while excluding the data gap due to Earth's occultation and the South Atlantic Anomaly (SAA). We used data from all layers of LAXPC 10 and LAXPC20 for the temporal analysis of GRS 1915 + 105.  The LAXPC 30 detector has shown gain change with time due to detector gas leakage, and hence, its data is not used. It is important to highlight that we exclusively extract the source spectra from LAXPC 20 data due to the consistent stability of its gain throughout the entire observational period, as previously reported by \citet{antia2021large}.

\subsection{SXT}
The Level 1 data obtained from the SXT instrument's photon counting mode underwent processing using the official SXT pipeline AS1SXTLevel2 - 1.4b\footnote{\href{https://www.tifr.res.in/~astrosat_sxt/sxtpipeline.html}{https://www.tifr.res.in/~astrosat\_sxt/sxtpipeline.html}} in order to generate Level 2 mode data. For the analysis of all observation sets listed in Table \ref{tab:table1}, the Photon Counting mode (PC mode) data were selected. The HEASoft tool XSELECT (version 6.29) was used to produce spectra, light curves, and images from SXT data. The response matrix file (RMF) \textit{sxt\_pc\_mat\_g0to12\_RM.rmf}, standard background spectrum \textit{SkyBkg\_Burst\_EL3p5\_Cl\_Rd16p0\_v01.pha} and ancillary response file (ARF) \textit{sxt\_pc\_excl00\_v04\_20190608\_mod\_16oct} \\
\textit{21.arf} were used for the analysis. Additionally, a correction for offset pointing was applied using the sxtARFmodule provided by the SXT instrument team.  During high flux states when average X-ray flux exceeds 40 c/s, the SXT flux is corrected for pile up. 

\begin{table*}[ht]
\centering
\scriptsize
\caption{Comprehensive details of the AstroSat observations of the source GRS 1915+105 during the period between Jul 2017 and Sep 2017. The table lists observation IDs alongside the corresponding exposure time, date, and time of each observation. \label{tab:table1}}
\begin{tabular}{lcccccc}
\hline
Observation  & Observation Date & Observation ID & Start Time (hh:mm:ss) & End Time (hh:mm:ss) & LAXPC & SXT \\ 
No. & & & & & Exposure & Exposure \\ 
& & & & & Time (ks) & Time (ks) \\
\hline
1 & 27 Jul 2017 (MJD 57961) & G07\_028T01\_9000001406 & 10:09:26 & 12:41:37 & 3.28 & 3.27 \\
2 & 27 Jul 2017 (MJD 57961) & G07\_046T01\_9000001408 & 14:18:25 & 04:45:32 & 19.69 & 10.96 \\
3 & 30 Aug 2017 (MJD 57995) & G07\_028T01\_9000001500 & 02:16:35 & 09:54:57 & 10.16 & 4.50 \\
4 & 31 Aug 2017 (MJD 57996) & G07\_046T01\_9000001506 & 11:11:25 & 00:03:20 & 10.47 & 15.31 \\
5 & 11 Sep 2017 (MJD 58007) & G07\_046T01\_9000001534 & 13:38:55 & 05:17:27 & 16.92 & 10.97 \\
\hline
\end{tabular}
\end{table*}

\begin{figure*}
     \subfloat{
         \includegraphics[width=0.49\textwidth]{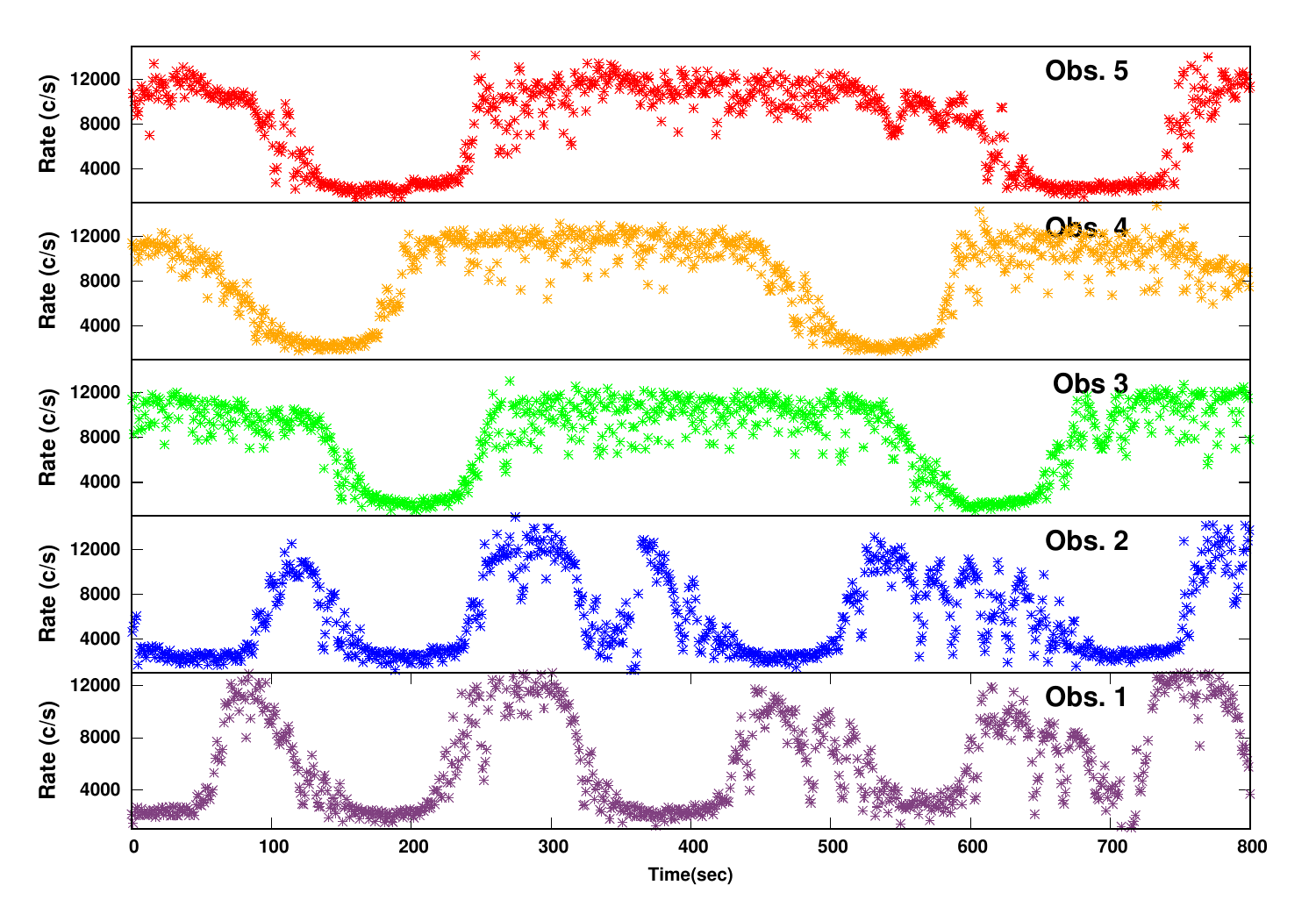}
     }
     \hfill
     \subfloat{
         \includegraphics[width=0.49\textwidth]{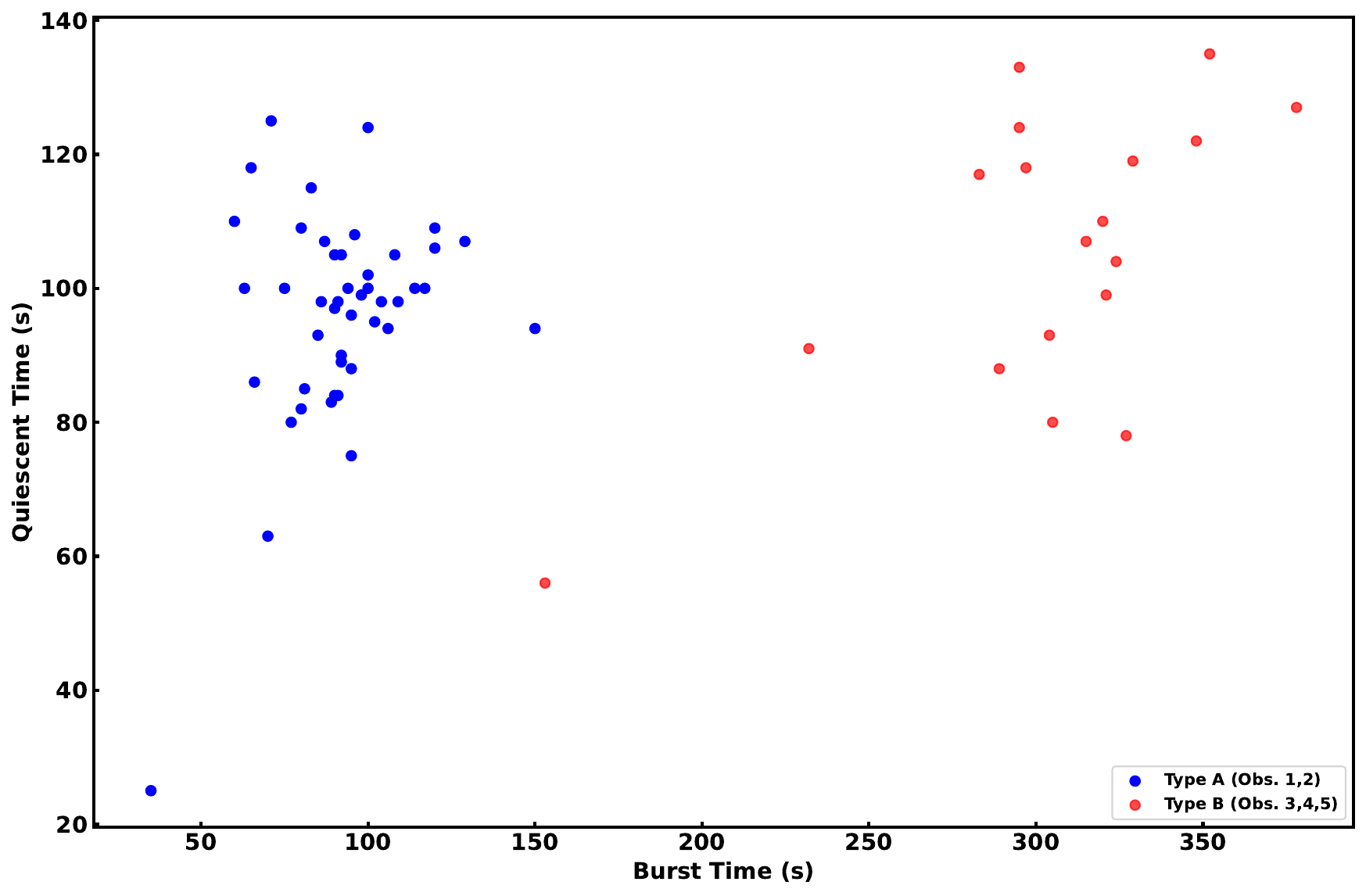}
     }
     \caption{Left panel: 800-sec background-subtracted light curves of GRS 1915+105 generated from all LAXPC detectors for Obs 1, Obs 2, Obs 3, Obs 4, and Obs 5 are shown in the different panels from top to bottom. For date of observation, see table \ref{tab:table1}. The time bin of all the light curves is 1 s. Right panel: plot of the burst duration vs the preceding quiescent time for type A bursts  (obs 1 \& 2) and for type B bursts (obs 3, 4, \& 5).}
\label{fig:fig 1}
\end{figure*}

\begin{figure*}
     \subfloat{
         \includegraphics[width=0.49\textwidth]{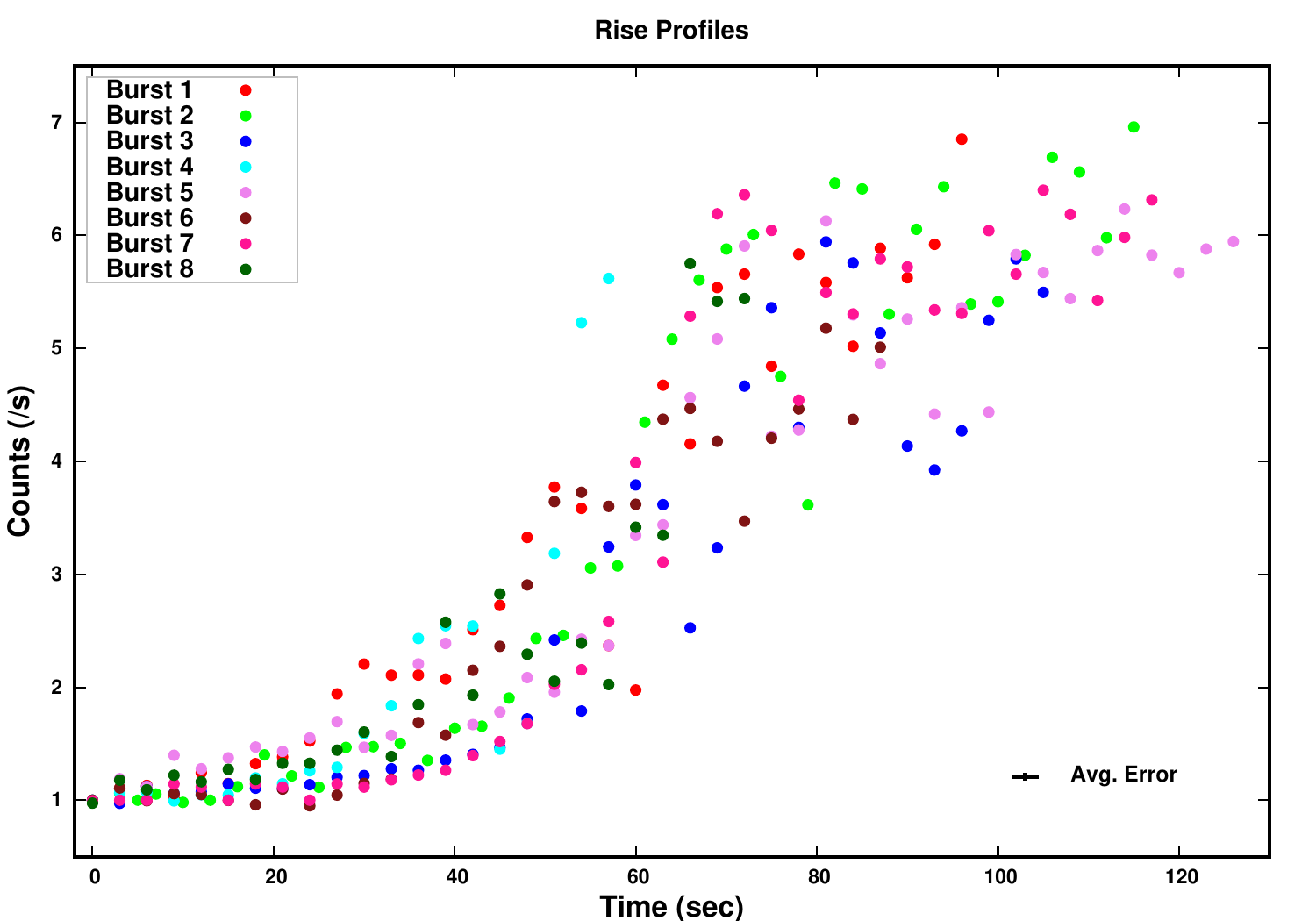}
     }
     \hfill
     \subfloat{
         \includegraphics[width=0.49\textwidth]{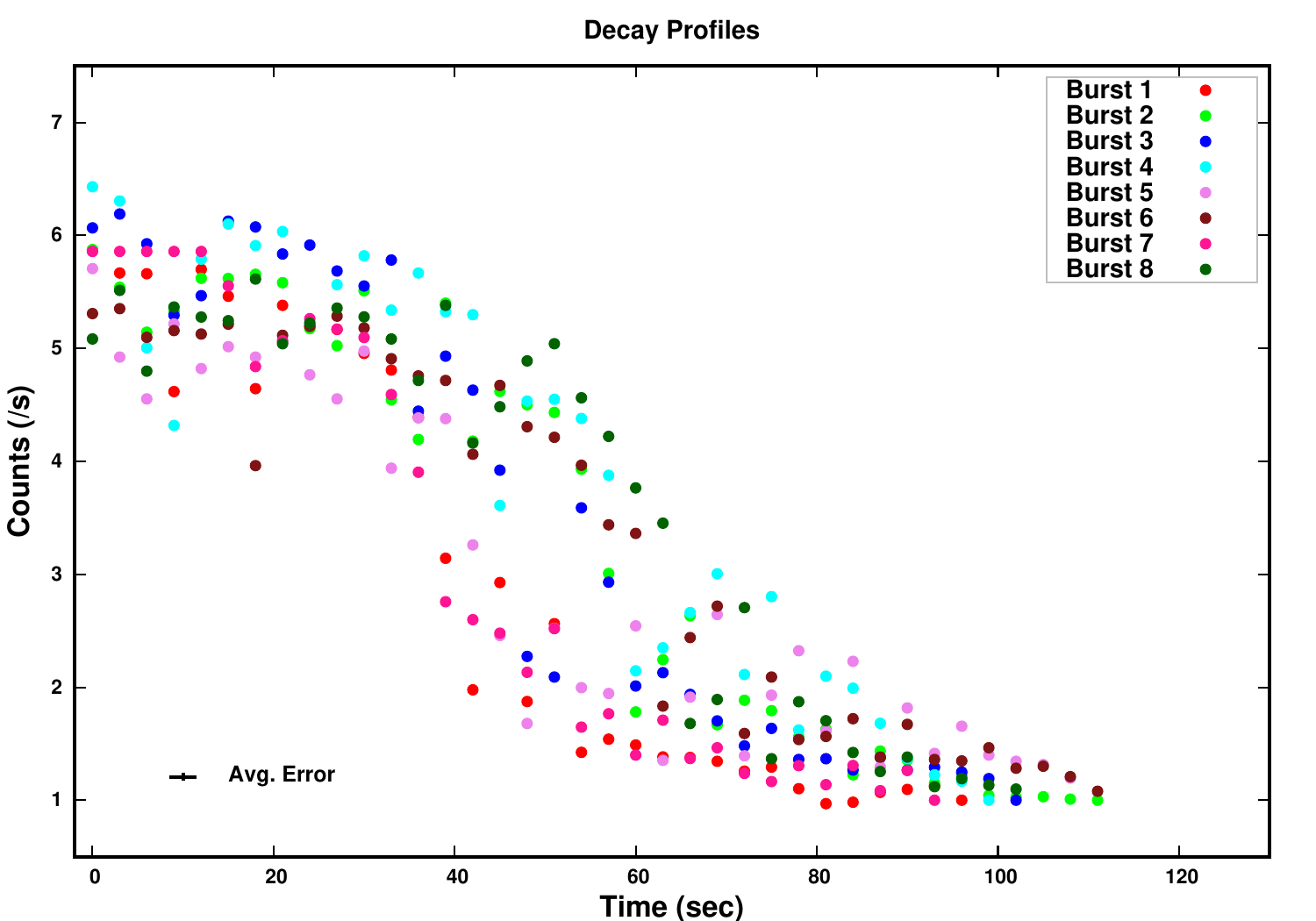}
     }
     \caption{Left Panel: Rise profile of $\eta$ flaring class of GRS 1915+105. The intensity of the rising bursts is normalized to the start point. Right panel:  Decay profile of $\eta$ flaring class. The intensity of the decay segment is normalized to the endpoint. The black point in both the figures represents the average error bars.}
\label{fig:fig 2}
\end{figure*}

\begin{figure*}
     \subfloat{
         \includegraphics[width=0.49\textwidth]{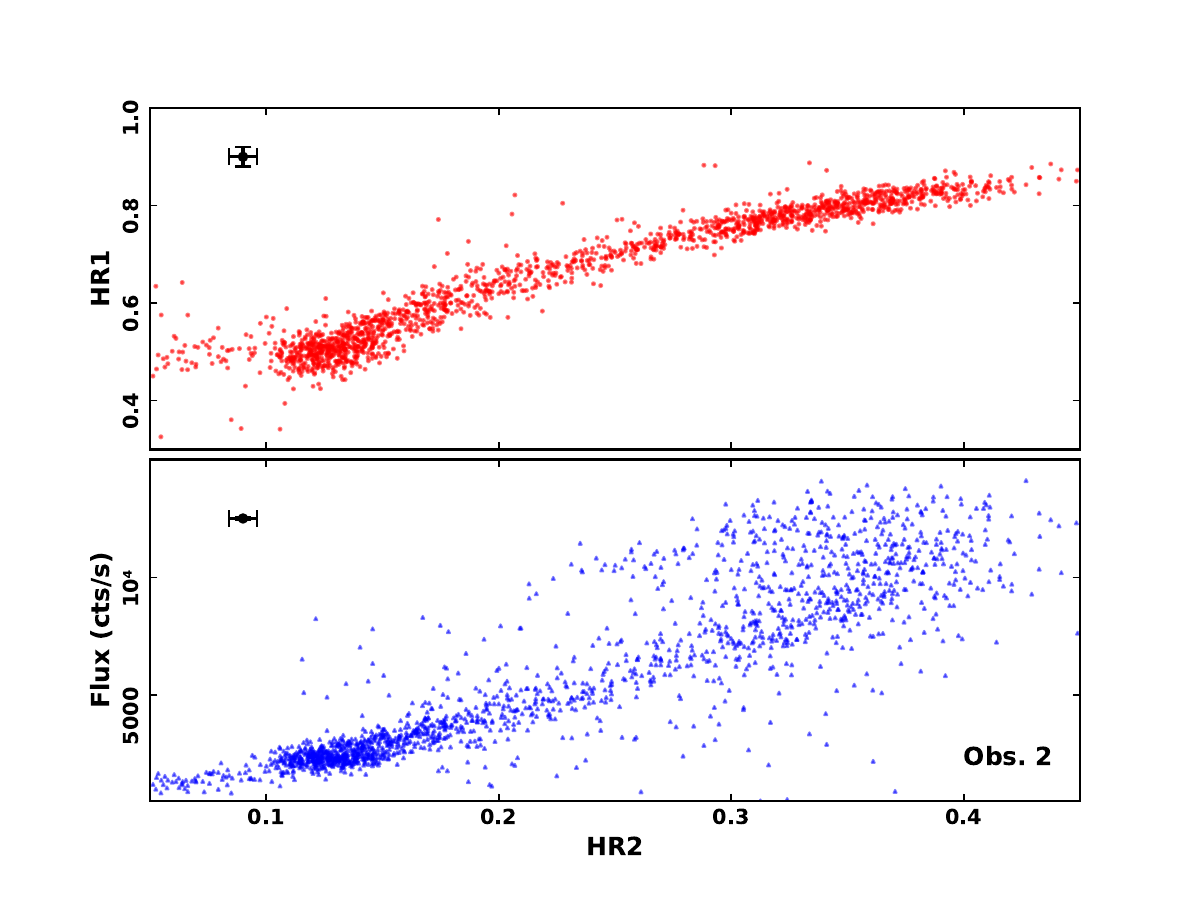}
     }
     \hfill
     \subfloat{
         \includegraphics[width=0.49\textwidth]{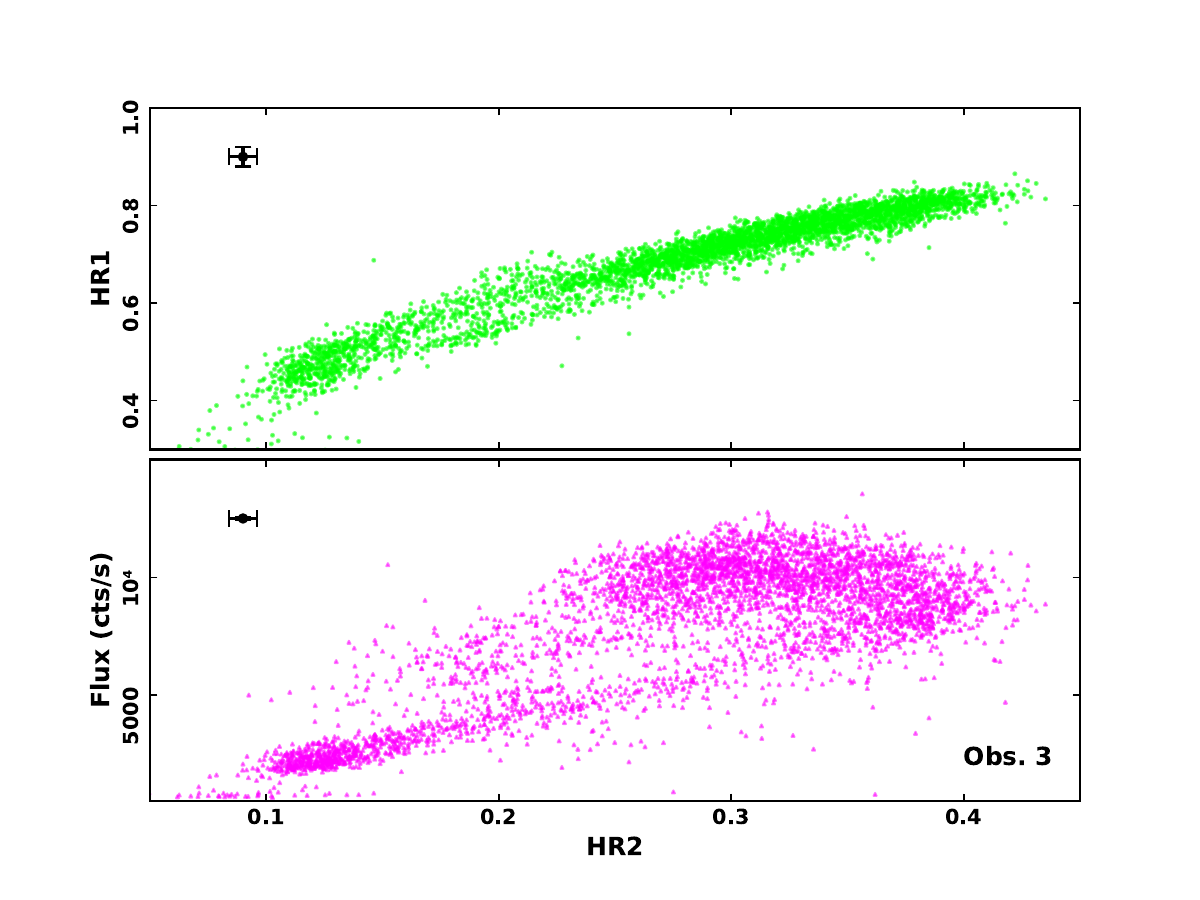}
     }
     \caption{Left panels and right panels show data of type A bursts (obs. 2) and type B bursts (obs. 3), respectively. The top panels show  CCD diagrams, while the bottom panels show  HID diagrams.  HR1 is taken as ratio of flux  of  6-10 keV energy range to  flux in 3-6 keV energy range while   HR2  is defined as ratio of flux in 10-30 keV  energy range to 3-6 keV energy range.) X-ray flux on the y-axis is in the energy range 3-30 keV. The black data point in each panel represents the typical measurement uncertainty, with error bars indicating standard deviations along both axes.}
\label{fig:fig 3}     
\end{figure*}

\begin{figure}
    \centering
	\includegraphics[width=1.\columnwidth]{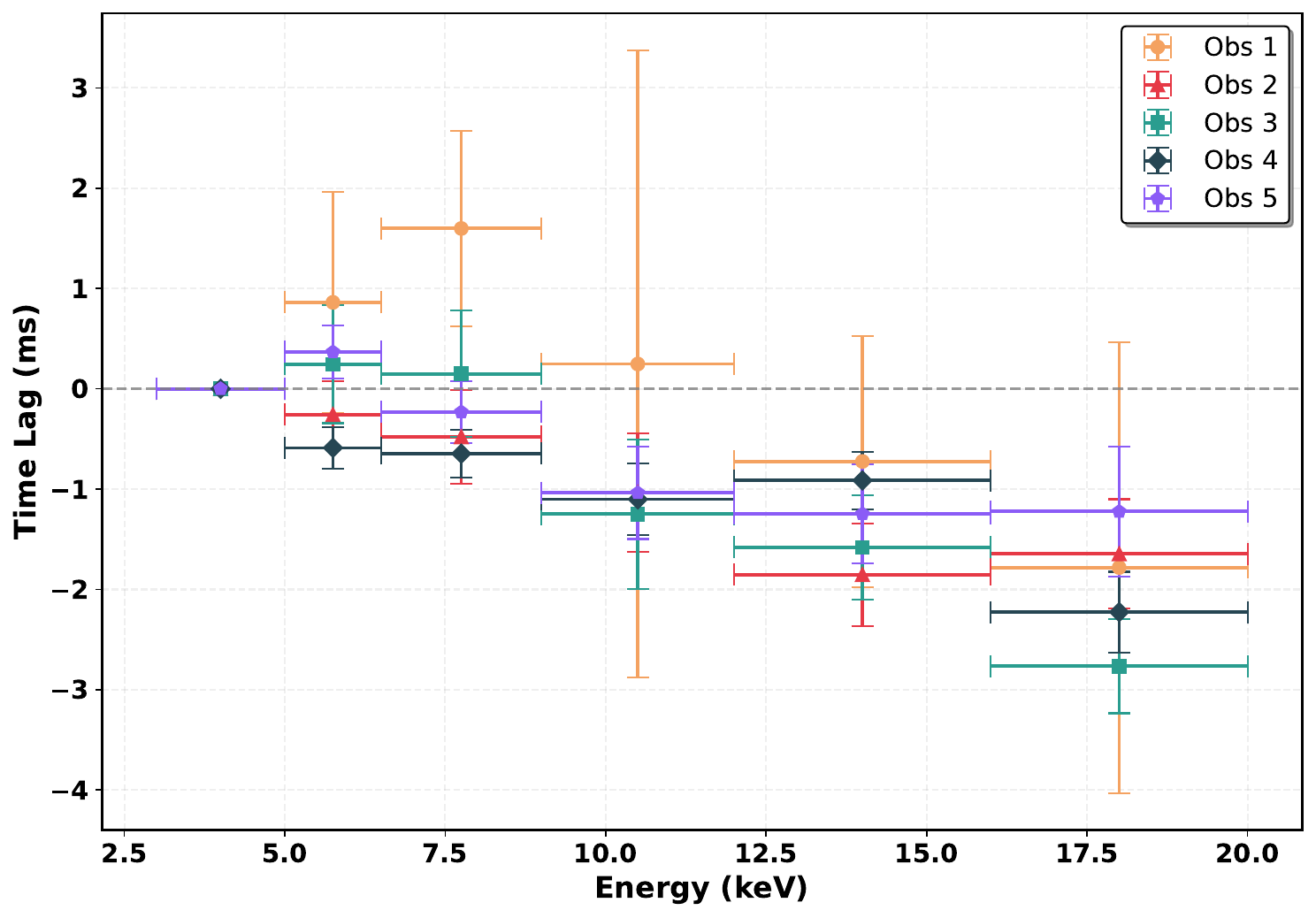}
        \caption{Energy-dependent time lag w.r.t. 3–6 keV band at the HFQPO frequency is plotted for the new variability class observations. See text for details.}
\label{fig:fig 4}
\end{figure}

\begin{figure*}
     \subfloat{
         \includegraphics[width=0.49\textwidth]{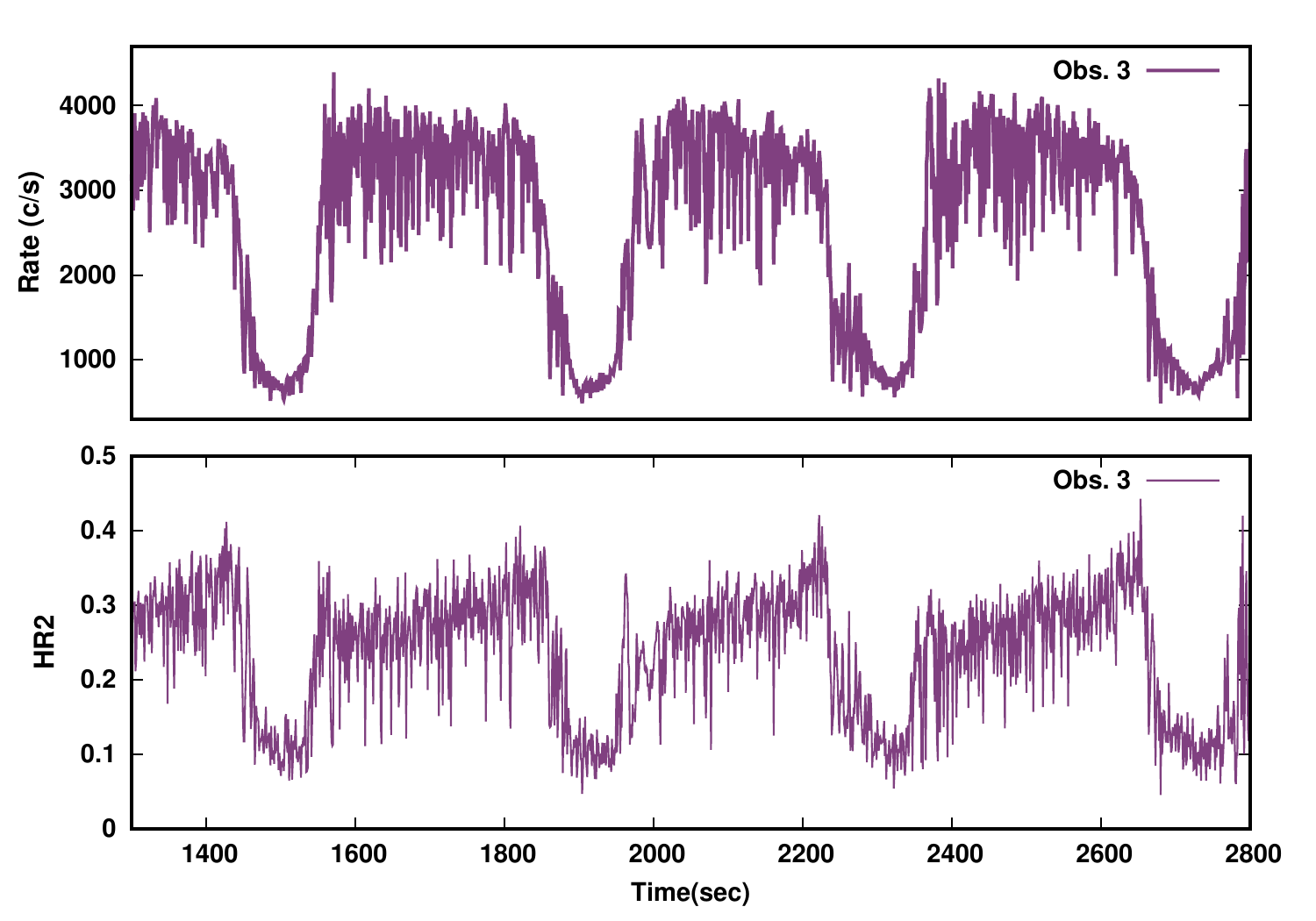}
     }
     \hfill
     \subfloat{
         \includegraphics[width=0.49\textwidth]{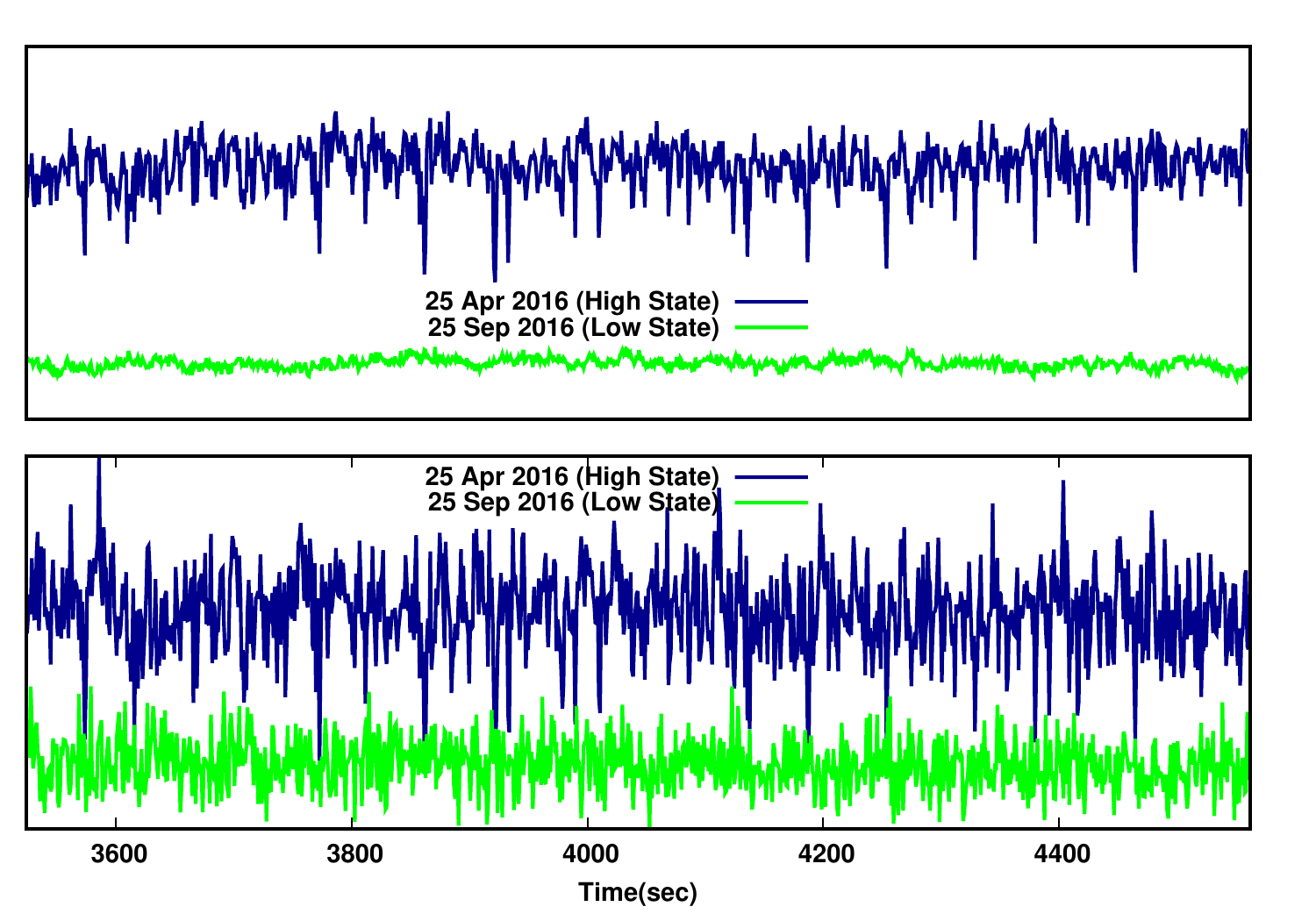}
     }
     \caption{Left top panel shows light curve for type B bursts (obs. 3) using 3-30 keV LAXPC 20 data.  The right top panel shows light curves of persistent high flux and persistent low flux Soft states using data from 25 Apr 2016 and 25 Sep 2016, respectively, for comparison with $\eta$ class.
        The left bottom panel shows  HR2 vs. time for type B bursts (obs. 3), and the right bottom panel shows   HR2 vs. time for the two persistent high flux and low flux Soft states. See text for more details.}
        \label{fig:fig 5}
\end{figure*}

\begin{table*}[ht]
\centering
\scriptsize
\caption{Details of the best-fitting PDS parameters from LAXPC 20 observations of GRS 1915+105 in the 6--20 keV energy band. 
\label{tab:table2}}
\begin{tabular}{lcccc}
\hline
Date & QPO Frequency (Hz) & Quality Factor & rms (\%)  & Significance    \\
\hline
\multicolumn{5}{c}{New Class $\eta$} \\
\hline
Obs. 1 (Flare high flux) & $70.73 \pm 0.31$ & $11.92 \pm 1.70$   & $3.56 \pm 0.21$ & 9.12      \\
Obs. 1 (Flare low flux)  & $69.92 \pm 0.98$ & $10.81 \pm 4.60$   & $4.24 \pm 0.67$ & 3.48      \\
Obs. 2 (Flare high flux) & $70.39 \pm 0.15$ & $14.25 \pm 1.12$   & $3.48 \pm 0.11$ & 16.62      \\
Obs. 2 (Flare low flux)  & $70.08 \pm 0.64$ & $12.62 \pm 3.66$   & $3.44 \pm 0.41$ & 4.46      \\
Obs. 3 (Flare high flux) & $68.29 \pm 0.17$ & $16.33 \pm 0.90$   & $3.33 \pm 0.12$ & 13.60      \\
Obs. 3 (Flare low flux)  & $69.78 \pm 0.37$ & $32.84 \pm 15.97$  & $3.11 \pm 0.59$ & 2.73     \\
Obs. 4 (Flare high flux) & $68.75 \pm 0.37$ & $14.12 \pm 0.68$   & $3.53 \pm 0.12$ & 15.11      \\
Obs. 4 (Flare low flux)  & $69.76 \pm 0.64$ & $21.71 \pm 9.69$   & $3.79 \pm 0.72$ & 2.82       \\
Obs. 5 (Flare high flux) & $71.37 \pm 0.14$ & $19.44 \pm 0.93$   & $2.92 \pm 0.12$ & 12.95      \\
Obs. 5 (Flare low flux)  & $71.04 \pm 1.45$ & $16.93 \pm 7.94$   & $3.36 \pm 1.15$ & 1.88       \\

\hline
\end{tabular}
\end{table*}

\begin{figure*}
     \centering
     \subfloat{
         \includegraphics[width=0.49\textwidth]{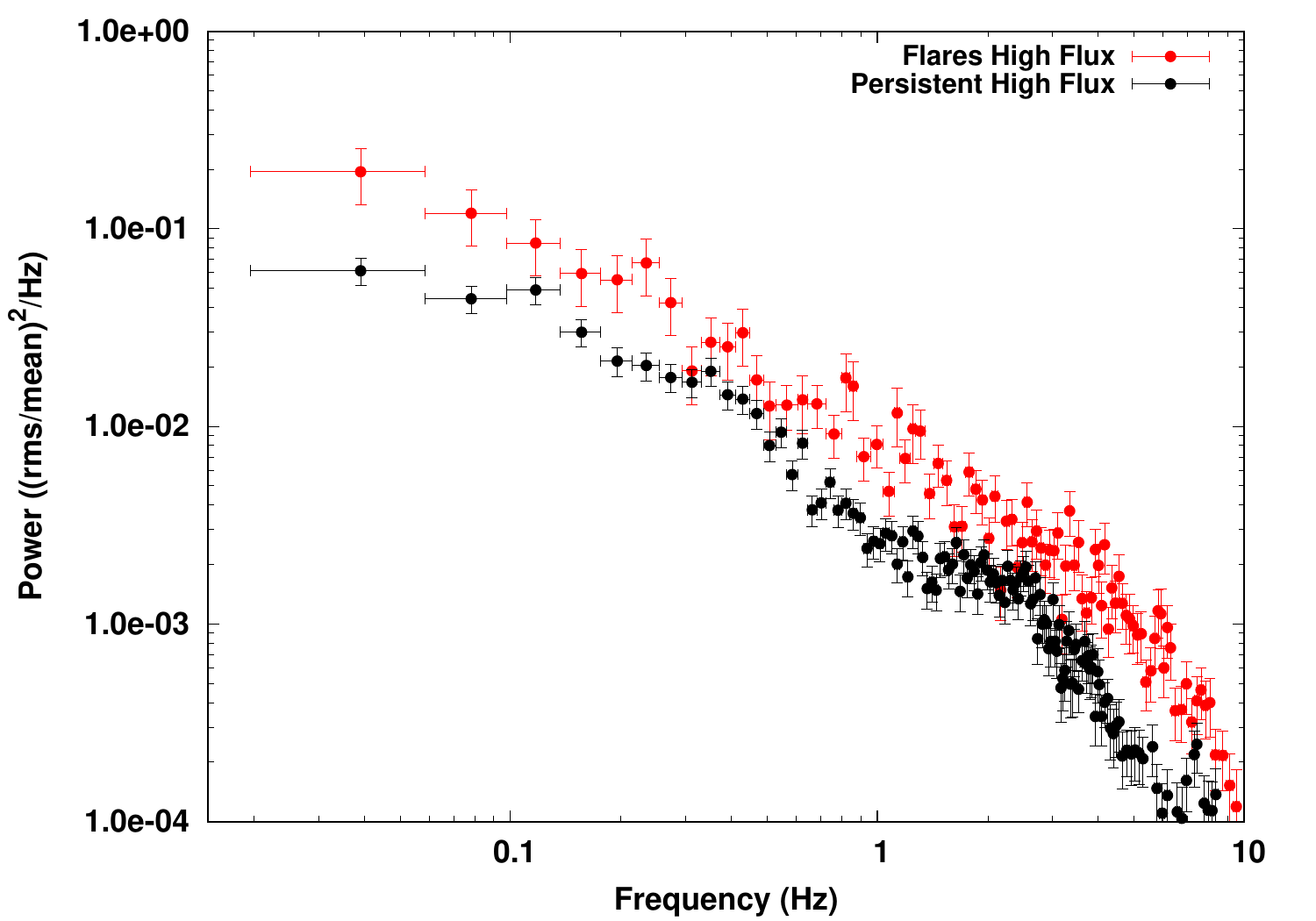}
     }\subfloat{
         \includegraphics[width=0.49\textwidth]{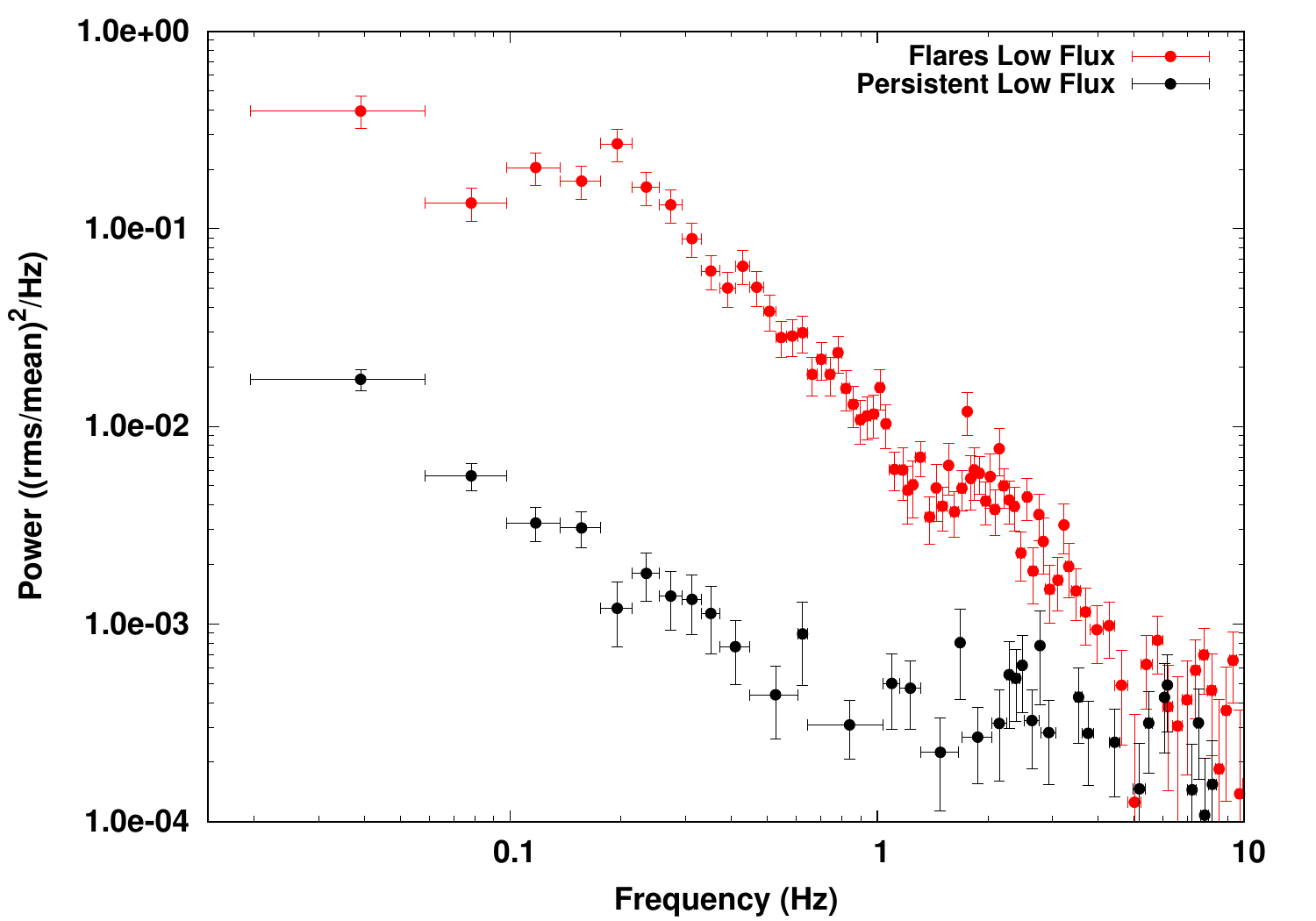}
     }
     \par\medskip
     \subfloat{
         \includegraphics[width=0.49\textwidth]{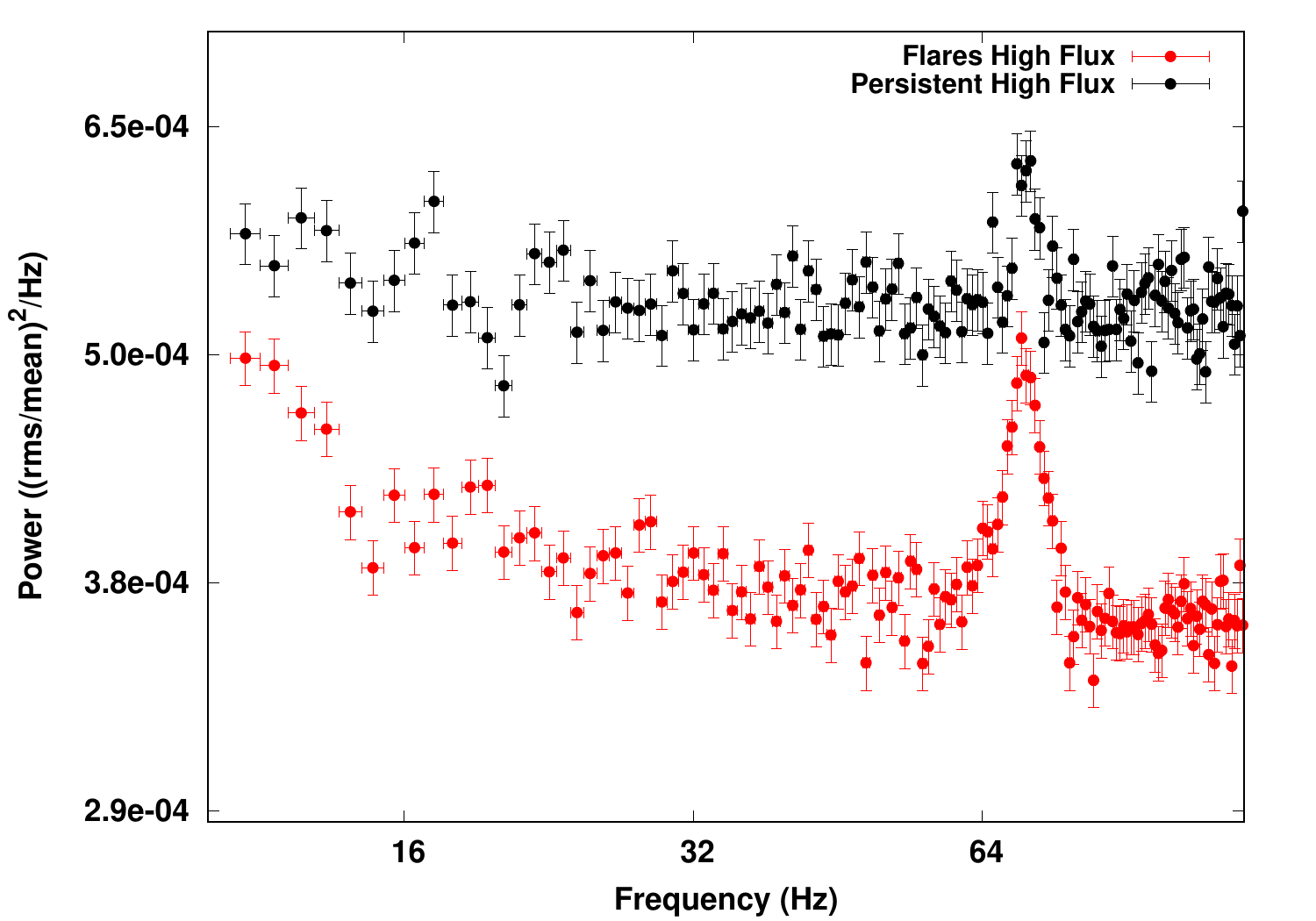}
     }
     \subfloat{
         \includegraphics[width=0.49\textwidth]{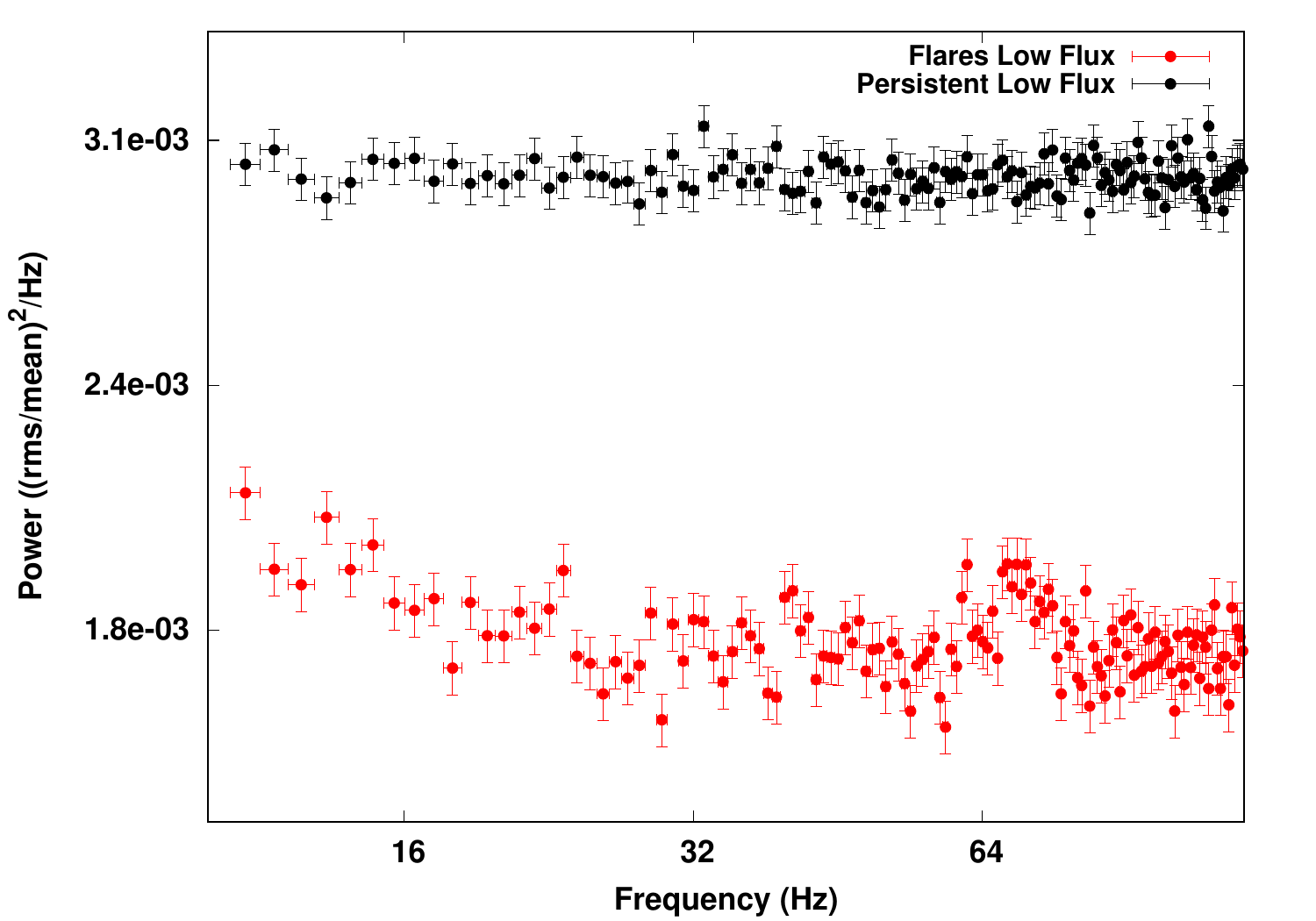}
     }
    \caption{Left panels show the power density spectra (PDS) of GRS 1915+105 for the persistent high flux (in black)  and flare high flux segment (Obs. 5, in red) across two frequency ranges: 0.05–10 Hz (upper-left) and 10–120 Hz (lower-left) using LAXPC 10 and 20 data. The right panels present the PDS for the persistent low flux  (in black) and flare low flux segment (Obs. 5, in red) in the two frequency ranges as above. In the low-frequency range (top panels), the PDS are Poisson noise corrected, whereas the high-frequency PDS (bottom panels) are not corrected.  See text for details.}
        
        \label{fig:fig 6}
\end{figure*}

\begin{figure*}
     \subfloat{
         \includegraphics[width=0.49\textwidth]{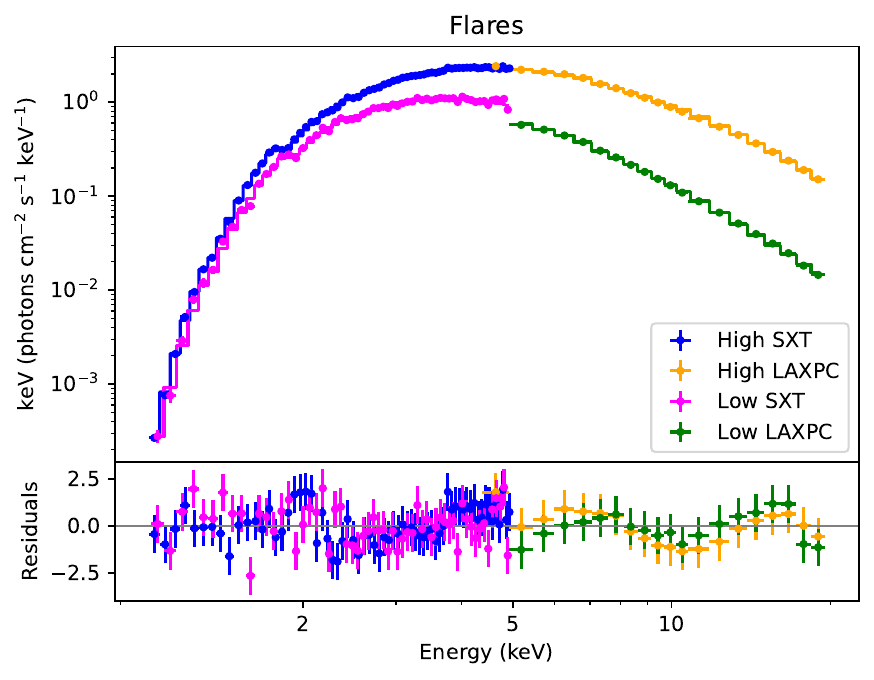}
     }
     \hfill
     \subfloat{
         \includegraphics[width=0.49\textwidth]{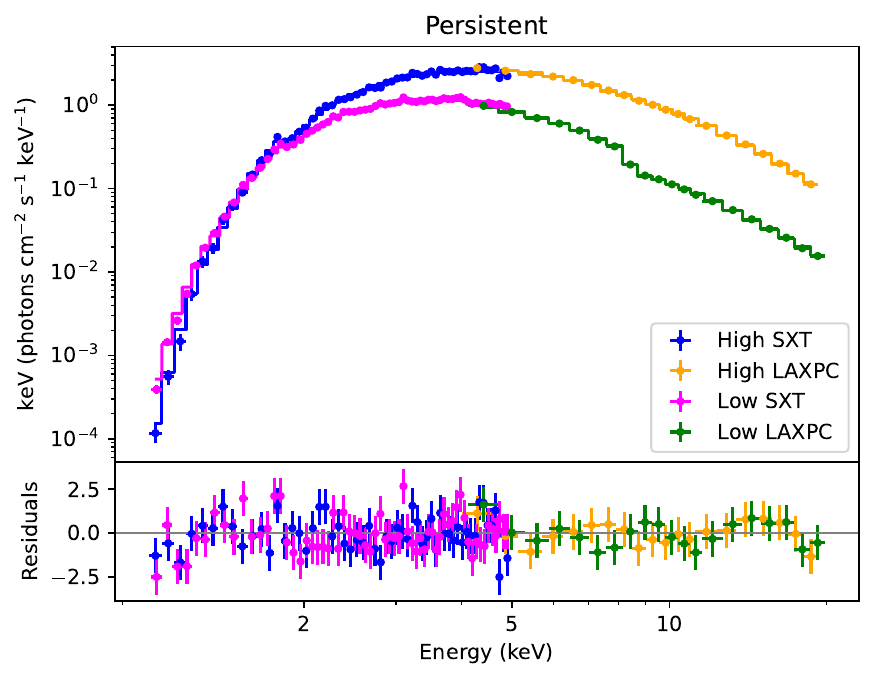}
     }
     \caption{Left panel shows energy spectra of flares ($\eta$ X-ray flaring class, obs. 3) for high flux and low flux segments separately while right panel shows energy spectra of persistent high flux and persistent low flux  X-ray states (refer Figure \ref{fig:fig 5}).  The same color scheme is used for both panels: High flux;   SXT  in blue, \& LAXPC 20  in orange, and for low flux, SXT  in magenta \&  LAXPC 20  in green. See text for further details.}
\label{fig:fig 7}
\end{figure*}

\begin{table*}[ht]
\centering
\scriptsize
\caption{Best-fit spectral parameters of GRS 1915+105. \label{tab:table3}}
\begin{tabular}{lcccccccc}
\hline
Date & $n_H$ & Photon & Fraction & Accretion & Inner & $\chi^2/\mathrm{bins}$ & $\chi^2/\mathrm{bins}$ & $\chi^2/\mathrm{dof}$ \\
     & ($10^{22} \mathrm{cm}^{-2}$) & Index ($\Gamma$) & Scatter & Rate ($10^{18} \mathrm{g/s}$) & Radius ($R_g$) & (SXT) & (LAXPC20) & (Total) \\
\hline

Obs. 1 (Flare high flux)    & $4.24 \pm 0.03$  & $4.6 \pm 0.1$ & $0.89 \pm 0.02$ & $1.8 \pm 0.1$ & $< 1.6$ & 75.3/59  & 9.0/20  & - \\
Obs. 1 (Flare low flux)                            & $tied$           & $4.40 \pm 0.03$ & $0.50 \pm 0.01$ & $0.57 \pm 0.01$ & $2.18 \pm 0.03$ & 43.7/58  & 8.7/19  & 136.7/141 \\
Obs. 2 (Flare high flux)    & $4.40 \pm 0.02$  & $>4.8$          & $>0.99$         & $1.88 \pm 0.03$ & $1.76 \pm 0.07$ & 112.1/66 & 31.4/20 & - \\
Obs. 2 (Flare low flux)           & $tied$           & $4.39 \pm 0.03$ & $0.48 \pm 0.01$ & $0.59 \pm 0.01$ & $2.13 \pm 0.03$ & 90.3/67  & 7.9/19  & 241.6/157 \\
Obs. 3 (Flare high flux)    & $4.27 \pm 0.04$  & $4.5 \pm 0.2$ & $0.8 \pm 0.2$ & $1.46 \pm 0.08$ & $<2.1$          & 56.1/67  & 12.7/20 & - \\
Obs. 3 (Flare low flux)     & $tied$           & $4.4 \pm 0.1$ & $0.54 \pm 0.07$ & $0.55 \pm 0.03$ & $2.4 \pm 0.3$ & 59.7/59  & 8.9/19  & 137.6/150 \\
Obs. 4 (Flare high flux)    & $4.25 \pm 0.02$  & $4.64 \pm 0.10$ & $>0.98$         & $1.46 \pm 0.01$ & $<1.6$          & 94.8/69  & 13.5/20 & - \\
Obs. 4 (Flare low flux)    & $tied$           & $4.46 \pm 0.03$ & $0.58 \pm 0.01$ & $0.58 \pm 0.01$ & $2.59 \pm 0.03$ & 42.0/60  & 13.3/19 & 163.6/153 \\
Obs. 5 (Flare high flux)    & $4.26 \pm 0.02$  & $>4.59$         & $>0.98$         & $1.5 \pm 0.1$ & $<1.7$          & 49.8/71  & 15.9/20 & - \\
Obs. 5 (Flare low flux)  & $tied$           & $4.21 \pm 0.04$ & $0.37 \pm 0.01$ & $0.68 \pm 0.01$ & $2.91 \pm 0.03$ & 40.9/59  & 3.6/19  & 110.2/154 \\
Full flare event (Obs. 2)   & $4.35 \pm 0.04$  & $4.35 \pm 0.06$ & $>0.95$         & $0.89 \pm 0.02$ & $<1.8$          & 79.1/72  & 8.2/20  & 87.2/84   \\
Full flare event (Obs. 5)   & $4.39 \pm 0.01$  & $4.83 \pm 0.07$ & $>0.99$         & $1.38 \pm 0.01$ & $<1.5$          & 60.80/71 & 24.6/20 & 85.4/83  \\
Persistent high flux        & $4.6 \pm 0.1$  & $4.8 \pm 0.4$   & $0.7 \pm 0.4$   & $2.3 \pm 0.4$   & $2.9 \pm 0.5$   & 46.7/57  & 11.7/20 & 58.4/68   \\
Persistent low flux         & $3.83 \pm 0.09$  & $4.4 \pm 0.4$   & $0.4 \pm 0.1$   & $0.8 \pm 0.1$   & $<3.2$          & 66.4/64  & 10.0/20 & 76.5/76   \\
\hline
\end{tabular}
\end{table*}

\section{Results}
\subsection{New flaring variability class}
\label{sec:the_new_var_classes}
We have generated background-subtracted light curves for all observations carried out during  July - September 2017 (refer to Table \ref{tab:table1})  using data from all three LAXPC units 
with a time bin of 1.0 seconds. The left panel of Figure \ref{fig:fig 1} shows an 800-second background-subtracted light curves in the energy range 3-30 keV. Each panel represents the light curves from Observations 1 to 5, arranged sequentially from bottom to top. Notably, these observations exhibit oscillations in count rate, with the higher count rate (burst phase)  being approximately 6-7 times higher than the lower count rate (quiescent phase), which falls in higher ratio region \citep[See figure 8 of][]{yadav1999different}. 
We define an event as a full burst cycle (burst period P)  having a quiescent phase  (low flux segment) followed by a burst phase (high flux segment). We measured the start of a burst event corresponding to the end of the previous burst event.
With this definition, there is no gap between burst events \citep[][]{belloni1997unified, yadav1999different}. Thus, three quantities are associated with a burst event: the burst cycle recurrence time (burst period P), the duration of the quiescent phase, and the duration of the burst phase.  We separate the quiescent phase and burst phase by taking the separation line at flux  5000 c/s for all three LAXPC units; below this line is taken as the quiescent phase, and above this line is the burst phase. 

All the bursts shown in the left panel of figure \ref{fig:fig 1}, can broadly be put into two types:  Type A)  the mean recurrence time of the burst cycle is around 181 s with dispersion $\delta P/P \sim 22\% $ for observations 1 \& 2. These are quasi-regular bursts \citep[][]{yadav1999different}. In this type, the count rates during the burst phase reach up to 14000 c/s, and in type B)  when the flux is lower, as in observations 3, 4 \& 5, the source exhibits more regular bursts, and the mean recurrence of the burst cycle is around 392 s with dispersion  $\delta P/P \sim 13\% $. 
These bursts also fall under the quasi-regular type.
In type B bursts, the count rate during the burst phase reaches up to 12000 c/s. Both types of bursts show the presence of dips (quiescent phase segment) lasting around 100 s. All bursts show a slow rise and a slow decay. The shorter A-type bursts show oscillations toward the end of the burst cycle. A total of 48 quasi-regular A-type bursts with a recurrence time of 191 sec (observation 1  \& 2 ) and a total of 18 B-type bursts with a recurrence time of 392 sec (obs. 3, 4, \& 5) have been detected. 
 
Both types of bursts discussed above are categorized as quasi-regular bursts. \citet{yadav1999different} have studied quasi-regular bursts, a subclass of $\kappa$ flaring class observed by IXAE on  19 and 21 June 1997, which shows a strong correlation between quiescent duration and the following burst duration. The right panel of Figure \ref{fig:fig 1} shows a plot of the burst duration vs the preceding quiescent time for type A and type B bursts. We do not find any correlation between the burst duration and the following quiescent duration as reported for $\kappa$ flaring class \citep [][]{yadav1999different}. 
In the case of $\kappa$  flaring class, burst duration for most bursts is $\le$ 80 s with very few bursts with duration above  80 s  \citep [See figure 8 of][]{yadav1999different}. In our case for type B bursts, most of the bursts have a burst duration of more than 250 s, while very few are below it. In the case of type A bursts, most of the bursts have a duration of more than 80 s while very few are below it  (see right panel of figure \ref{fig:fig 1}). The observed burst duration sets $\eta$ flaring class apart distinctly from the $\kappa$ flaring class.
 In the left panel of Figure \ref{fig:fig 2}, we present the 'rise' profile of this new flaring class. The flux is normalized to the starting point of the rising segment. The right panel of Figure \ref{fig:fig 2} displays the 'decay' profile, where the flux is normalized to the endpoint of the decay segment.  For $\eta$ flaring class, both the rise and decay time are approximately 50 s duration, whereas in all previously known bursts, there is a slow rise and fast decay, with both rise and decay duration less than 10 s \citep[][]{belloni1997unified, yadav1999different}. This extended transition duration of this new burst class sets them apart distinctly from all previously known flaring classes.

Background-subtracted light curves are generated with a time bin of 1.0 s in three energy bands a: 3-6 keV, b: 6-10 keV, and c: 10-30 keV.  During all our observations, the source remains in the soft state and has an insignificant contribution above 30 keV.   We have defined soft color (hardness ratio), HR1 = b/a, and hard color (hardness ratio), HR2 = c/a. 
The top left panel of Figure \ref{fig:fig 3} shows the Color-Color Diagram(CCD) for observation 2 (type A), illustrating the variation of the hard color (HR2) with the soft color (HR1). As burst evolves from lower flux (quiescent phase) to higher flux (burst phase), HR2 increases from 0.1 to 0.4 while HR1 varies from 0.5 to 0.9.  In the bottom left panel, a hardness-intensity diagram (HID) is presented, illustrating the variation of total flux extracted in the energy band of 3-30 keV with HR2. X-ray flux varies from  1000 c/s to almost 14000 c/s.  In the top right panel, the CCD diagram for Observation 3 (type B) is shown, demonstrating the variation of HR1 with HR2. Here, as the burst evolves from lower flux to higher flux, HR2 increases from 0.1 to 0.4 (same as type A). Source spends more time in a high flux state in this type, resulting in a cluster at higher HR1 \& HR2 values, which is different from the results for type A. In the bottom right panel, a CCD diagram is presented for type B, illustrating the variation of total flux with HR2. The source count rate varies from  1000 c/s  to around 12000 c/s (lower than type A bursts). In both types of bursts, the source undergoes the transition from low flux to high flux, and HR2 also increases from 0.1 to 0.4. \citet{belloni2000model} have studied the CCD diagram of $\kappa$, $\lambda$ and $\rho$ flaring classes observed in GRS 1915+105 using RXTE/PCA data. The source oscillates between the HIMS state and HS state and shows a cyclic pattern in the CC diagram. In this case, HR1 is defined as a ratio of  5-13 keV band and 2-5 keV band flux, while HR2  is defined as the ratio of X-ray flux in the 13-60 keV band and 2-5 keV band as per the energy range of RXTE/PCA (2--60 keV). In our case, it is almost a linear transition and no significant cyclic pattern in the CCD.  The $\omega$ flaring class also shows similar behavior \citep[][]{hannikainen2005characterizing}.  It is worth noting here that the effective area of   LAXPC detectors is at least 3 times higher than RXTE/PCA above 13 keV, and so, LAXPC, along with SXT, can easily distinguish two HS states, as will be clear from our energy spectral analysis results presented in the next section \citep[][]{yadav2016large}.  

\subsection{Timing Characteristics}
To study the timing characteristics of the new  $\eta$  class, we divide the burst cycle into a high flux zone and a low flux zone.  The high flux zone consists of the high flux segment (burst phase) when X-ray flux remains above 8000 c/s. The low flux zone consists of the low flux segment (quiescent/dip phase) when X-ray flux falls below 4000 c/s. These criteria is adopted to remove steep transition data from our study of the quiescent phase and the burst phase separately. A burst event is also referred as flare. Flare high flux and flare low flux correspond to the burst phase and quiescent phase, respectively. 

To analyze the variability within the LAXPC data, we separate the light curve into low and high flux segments and extract the PDS for each flux category independently. We employ the \texttt{laxpc\_find\_freqlag} subroutine for PDS computation.  The subroutine first generates light curves with a time resolution of \(1.24 \times 10^{-3}\) s. The light curves are then divided into multiple segments, each with a duration of 1.28 s (corresponding to 1024 bins per segment). The PDS is computed for each segment, with a Nyquist frequency of \(\sim400\) Hz. The individual PDS segments are subsequently averaged to obtain the final PDS. Additionally, the LAXPC code accounts for the background while generating the PDS.

The PDS in the 20-120 Hz was modeled using a power-law component, a constant for the Poisson level, and a single Lorentzian function to account for the HFQPO. All parameters were kept free for the high-flux PDS, while for the low-flux PDS, the energy and width of the Lorentzian were tied to the values from the high-flux PDS. The Lorentzian normalization, along with the rest of the parameters, was allowed to vary. To assess the presence of HFQPOs, we adopted the standard criteria: a quality factor ($Q = \text{LC}/\text{LW} \geq 3$) and significance ($\sigma = \text{LN}/\text{err}_{\text{neg}} \geq 3$), where LC, LW, LN, and $err_{neg}$ correspond to the centroid frequency, width, normalization, and negative error of the Lorentzian normalization, respectively. Additionally, the root mean square (RMS) of the HFQPO was computed as $\sqrt{\text{LN}}$.  Table 2 summarizes the HFQPO characteristics in 6-20 keV energy band, including QPO frequency, quality factor, RMS, and significance. We detected HFQPOs in the 68.29–71.37 Hz range, with significance values spanning 1.88$\sigma$ to 16.62$\sigma$ and an rms\% between 2.92\% and 4.24\%. Importantly, the HFQPO rms\% did not show significant variation between low and high flux segments, suggesting that HFQPOs persist regardless of flux level. Our findings indicate that HFQPOs were detected in the 6–20 keV range, while no significant features were observed in the 3–6 keV and 20–60 keV PDS. The rms\% values in these energy bands were statistically insignificant, and the data quality is not sufficient to make any conclusive statements about the presence or absence of HFQPOs.

The LAXPC software task \texttt{laxpc\_find\_freqlag} computes energy-dependent time lags using the cross-correlation function, relative to a selected reference energy band over a specified frequency range ($\Delta f$) \citep[][]{2024ApJ...974...90D}. Figure \ref{fig:fig 4} illustrates the resulting time lags, with the reference energy band set to 3–6 keV. We observe that the soft time lag increases with energy, reaching approximately 1–2 ms at 15 keV.

\citet{yadav1999different} have studied  $\kappa$ and $\rho$ flaring classes as a transition between HS state and  HIMS state.  For a similar study, we use two high-state data of GRS 1915+105 observed on September 25, 2016 (low flux persistent  HS state), and on April 25, 2016 (high flux persistent  HS state).
 Figure \ref{fig:fig 5} presents the light curves and HR2 for new class $\eta$ and both the persistent soft states with  Low and High flux.  Background subtracted light curves are for the 3-30 keV band.  The  X-ray flux and HR2 of the burst phase of the $\eta$ class match well with the X-ray flux and HR2 of the persistent soft state with high flux.  Similarly, X-ray flux and HR2 of the low flux persistent HS state match well with  X-ray flux and HR2 of the quiescent phase (low flux segment of the class). 
 On 25 Apr 2016, GRS 1915+105 exhibited a high state with a persistent high flux level of approximately 3500 c/s, while on 25 Sep 2016, the source is in a steady low flux soft state at approximately 1000 c/s. 
 The HR2 stands at approximately 35 $\%$ ± 5$\%$  during the high flux HS state on 25 Apr 2016, whereas it decreases to 10$\%$ ± 2$\%$  during the low flux HS state on September 25, 2016.  
 It may be noted that the accretion disk is steady over an hour time scale during persistent HS states, while it is variable during the flaring bursts. We will discuss it further in the next section. \\

We present here our further study of GRS 1915+105, in particular timing characteristics of the flare low flux segment (quiescent phase) and the flare high flux segment (burst phase) separately and compare these with results for the two persistent HS states with low and high flux used to explain the transition during the bursts in Figure \ref{fig:fig 5}. PDS power of flare low and high flux segments and the persistent HS states crosses around 10 Hz, so we plot the data separately for low and high-frequency ranges. Our results are shown in Figure \ref{fig:fig 6}. Our PDS results for the high flux burst phase (flare high flux) and high flux persistent HS state are plotted in the left top panel in the 0.05--10 Hz frequency range and in the left bottom panel in the 10--120 Hz frequency range. Results for the flare low flux  (quiescent phase) and low flux persistent HS  state are plotted in the right top and right bottom panels in the frequency range 0.05--10 Hz and 10--120 Hz, respectively.  The  PDS power shows similar variation behavior for the flare high flux and the high flux HS persistent state below 10 Hz (left top panel), although the burst phase has more power than the high flux persistent HS state. However, this behavior is reversed above  10 Hz (left bottom panel) when the high flux persistent HS state has more power. HFQPO is clearly visible in both the flare burst phase (high flux) and the high flux persistent HS state. In the case of the quiescent phase (flare low flux) and the low flux persistent  HS state, there is a bigger difference in the frequency range below 10 Hz (top right panel), which is in contrast to high flux results (left top panel). Similar results can be seen in the higher frequency range (10--120 Hz) in the right bottom panel.  HFQPO is detected in the flare low flux (rms = 3.36 ± 1.15\%), but it is not significant in the low flux persistent HS state (rms <1\%).  The PDS power of flare dominates below 10 Hz, while the PDS power of persistent HS states dominates above  10 Hz. Overall, the timing behavior of the observed flares (bursts) is closer to the observed behavior of the high flux persistent HS state than that of the low flux persistent  HS state.
 
\subsection{Energy Spectral Analysis}
To study the energy spectrum, We have used the same criteria as used in PDS study for selecting the flare high flux   (burst phase) above 8000 c/s and flare low flux (quiescent phase) below 4000 c/s of the bursts. To construct the wide-band energy spectra, we combined data from the SXT and LAXPC20 instruments.  For the SXT, we considered the energy range of 1--5 keV, while for the LAXPC20, the energy range of 4--20 keV was used. Data below 1 keV is removed due to uncertainties, while data above 20 keV is ignored due to a low signal-to-noise ratio.
Absorption by the Inter-Stellar Medium (ISM) was taken into account with the TBabs model \citep{wilms2000absorption}  implemented with the galactic absorption abundance. The hydrogen column density for flare low flux spectra was tied with the flare high flux spectra. To consider the Comptonization of disk photons in the inner flow, we utilized the convolution model \textit{simpl} \citep[][]{steiner2009simple}{}{}. The inner radius of the disk and mass accretion rates were estimated from the best-fit values obtained from the relativistic disk model, \textit{kerrd} \citep[][]{ebisawa2003accretion}. We fixed the black hole mass, disk inclination angle, and distance to the source at 12.4M$_{\odot}$, 60 degrees, and 8.6 kpc, respectively \citep[][]{reid2014parallax}{}{}. The spectral hardening factor of \textit{kerrd} was also fixed to 1.7 \citep[][]{shimura1995spectral}. The left panel of Figure \ref{fig:fig 7} shows the energy spectra of  Observation 3  for both flare high flux and low flux segments. In this panel, the color blue corresponds to the energy spectrum obtained from the Soft X-ray Telescope (SXT), while the color orange represents the energy spectrum obtained from LAXPC 20 for the high flux level. The right panel of Figure \ref{fig:fig 7} shows the spectra of high and low flux persistent HS states for comparison. Overall, spectra of flare high flux and low flux (left panel) look similar to high and low flux persistent  HS states, respectively (right panel). 

In Table 
 \ref{tab:table3}, we present spectral analysis results for all observations, which include absorption column density, inner disk radius, accretion rate, scattered fraction, and photon index ($\Gamma$). We have analyzed flare low flux, flare high flux, total burst cycle, and the two persistent HS states separately. For all the observations, the photon index ($\Gamma$) is greater than 4, and the accretion inner radius
is less than 3.2 $R_g$, which suggests that the source was in the HS states during these observations. Therefore, the source oscillates between two persistent HS states during the $\eta$ flaring class. The accretion rate during the flare low flux  (quiescent phase) is in the range 0.55--0.68 $\times 10^{18}$ gm/s   while it increases to 1.46--1.88 $\times 10^{18}$ gm/s during the flare high flux (burst phase). The average accretion rate of a burst event (full burst) falls between these two values as expected. Accretion rates are  0.79 $\times 10^{18}$ gm/s  and 2.32 $\times 10^{18}$ gm/s during the low flux and the high flex persistent HS states, respectively. During type A $\eta$ bursts, the accretion rate during the flare high flux is higher than that for the flare high flux in type B bursts as expected from measured X-ray flux.

\section{Discussion}
\label{sec:discu}

As discussed earlier, the rise and decay time of these bursts is around 50 s. 
We can write the viscous timescale where $\alpha$ is the viscosity parameter. Where $\dot{m}_d$ is in the units of Eddington accretion rate, $m$ is the mass of the black hole in solar mass units, and $R_o$ is the inner disk radius 
in km. Then viscous timescale of the disk may be written as  \citep[][]{frank2002accretion}:
\begin{equation}
    t_{\mathrm{vis}}^d = 4.3 \times 10^{-4} \, \alpha^{-1} \dot{m}_d^{-1} m^{-1} R_o^2 \, \mathrm{s}
\end{equation}

For rise time, substituting $\dot{m}_d$=0.2, $m$=12.5, $R_o$=3 $\times R_g$=3 $\times$ 18.75 km and $\alpha$=0.01 and we get $t_{\mathrm{vis}}^d$   $\sim$ 54 s. This is close to the rise time observed in the $\eta$ class.  For decay time, substituting $\dot{m}_d$= 0.6, $m$=12.5, $R_o$= 2 $\times R_g$= 2 $\times$ 18.75 km and $\alpha$=0.01 and we get $t_{\mathrm{vis}}^d$= 8 s. This is also consistent with the observed rise time for given uncertainty in $\alpha$ and other parameters.  Here we take values of $R_o$ from the spectral analysis. The values of $\dot{m}_d$ are derived using the mass accretion rate estimates obtained from Table \ref{tab:table3}. The accretion rate increases by a factor of 3 when the source transitions from the low flux (dips) to the high flux (burst phase).

\citet{belloni2000model} have suggested that all the flaring classes observed in GRS 1915+105 can be explained in terms of transitions between states A, B, and C. The radio-quiet classes like $\lambda$, $\kappa$, $\rho$ and $\omega$ show mostly transition between B and C.  Radio loud flaring classes  $\beta$ and $\theta$  show transitions in all three states A, B, and C.  Transition from state B to state  C  produces hard dips which is similar as seen in the radio-quiet flaring classes. All these transitions are fast in less than 10 s.  However, when the source moves from state C to state B during $\beta$ or $\theta$ flaring class, it produces a spike, and the source transits to state A (very fast transition in 1-2 s) \citep[][]{yadav2001disk}. New flaring class $\eta$ can be explained as transitions between states B and A with  $\sim$ 50 s transition time. This class has only soft dips (no hard dips). Our results clearly show that new $\eta$ flaring class is very different from $\kappa$ and $\omega$ flaring classes in terms of burst profile, slow transition, no hard dips, burst phase duration, X-ray flux and HR2 in phase and the presence of HFQPO \citep[][]{yadav1999different, naik2002, belloni2019variable, athulya2022, majumder2022wide}. During our observations discussed here (MJD 57960-58006), GRS 1915+105 was in radio-quiet state as per available radio data \citep[][]{2021MNRAS.503..152M} and new $\eta$ class is likely to be a radio quiet flaring class.

We have analyzed Astrosat/LAXPC \& SXT data and report here a new flaring class, which we named $\eta$  class when the source oscillates between two HS states (the power index is always greater than 4) with transition time around  50 s. The X-ray flux is in phase with HR2. This class is quasi-regular, and we have detected $\sim$ 70Hz HFQPO  with soft lag. The accretion rate changes by a factor of almost three over the burst cycle. These bursts show the dependence of the burst duration with the preceding quiescent duration.  GRS~1915+105 has shown many flaring classes discovered earlier when the source oscillates between the High Soft state (HS) and the Hard Intermediate state (HIMS) with a transition time of less than 10 s. The  X-ray flux is anti-correlated with HR2  during all these flaring classes. HFQPO is never observed during these flaring classes.

\section*{Acknowledgements}
\label{sec:ack}

We thank the referee for the useful comments, which have improved the manuscript significantly. This work used data from the Soft X-ray Telescope (SXT) developed at TIFR Mumbai. And the SXT POC at TIFR is acknowledged for verifying and releasing the data through the Indian Space Science Data Centre (ISSDC) and providing the required software tools. We would also like to thank the LAXPC POC and SXT POC teams for their support. In addition,  this research has used the software provided by the High Energy Astrophysics Science Archive Research Center (HEASARC), a service of the Astrophysics Science Division at NASA.

\section*{Data Availability}
The observational data used in this paper are publicly available at ISRO’s Science Data Archive for AstroSat Mission (\url{https://astrobrowse.issdc.gov.in/astro_archive/archive/Home.jsp}) and  NASA’s High Energy Astrophysics Science Archive Research Center (HEASARC; \\ (\url{https://heasarc. gsfc.nasa.gov/})) and references are mentioned.

\bibliographystyle{aasjournal}

\end{document}